\documentclass[aps,amssymb,floatfix,showpacs]{revtex4}
\usepackage{graphicx}
\usepackage{epsfig}
\begin{document}

\title{More on the bending of light !}
\author{Kayll Lake \cite{email}}

\affiliation{Department of Physics, Queen's University, Kingston,
Ontario, Canada, K7L 3N6 }
\date{\today}
$\;$
\begin{abstract}
Recently, Rindler and Ishak have argued that the bending of light is, in principle, changed by the presence of a cosmological constant since one must consider not only  the null geodesic equation, but also the process of measurement. I agree with the fact that both must be considered. Here, on the basis of the mathematically exact solution to the classical bending problem, and independent of the cosmological constant, I clarify the approximate argument found in the vast majority of texts (new and old) for the measured value of the bending of light for a single source and show that the result is in part due to the almost perfect cancelation of two terms, one of which is seldom considered. When one considers two sources, this cancelation is of no consequence, and, for example, if the sources are opposite with the same associated apsidal distance, the approximate argument gives the rigorously correct answer (up to numerical evaluation), an answer which is unaffected by the presence of a cosmological constant.
\end{abstract}

\maketitle

\bigskip

\section{Introduction}
Recently, Rindler and Ishak  \cite{rindler} have corrected a long-standing error in the literature concerning the cosmological constant ($\Lambda$) and the bending of light for a single source, an error perpetuated by the author \cite{lake}. Here I reexamine the classic subject of the bending of light (with a concentration on a solar mass deflector - not gravitational lensing in general) and arrive at some further results which should be of wide - spread interest. The background geometry is the spherical vacuum given by
\begin{equation}
\label{vacuum}
ds^2 = \frac{d\textsf{r}^2}{f(\textsf{r})} + \textsf{r}^2(d\theta^2 + \sin^2 \theta  d\phi^2)-f(\textsf{r})dt^2, \label{line}
\end{equation}
where
\begin{equation}
\label{f}
f(\textsf{r}) = 1-\frac{2 m}{\textsf{r}}-\frac{\Lambda \textsf{r}^2}{3} \label{factor}
\end{equation}
and we restrict our analysis to $\theta=\pi/2$.

\bigskip

\section{Null Geodesics}
\begin{figure}[ht]
\epsfig{file=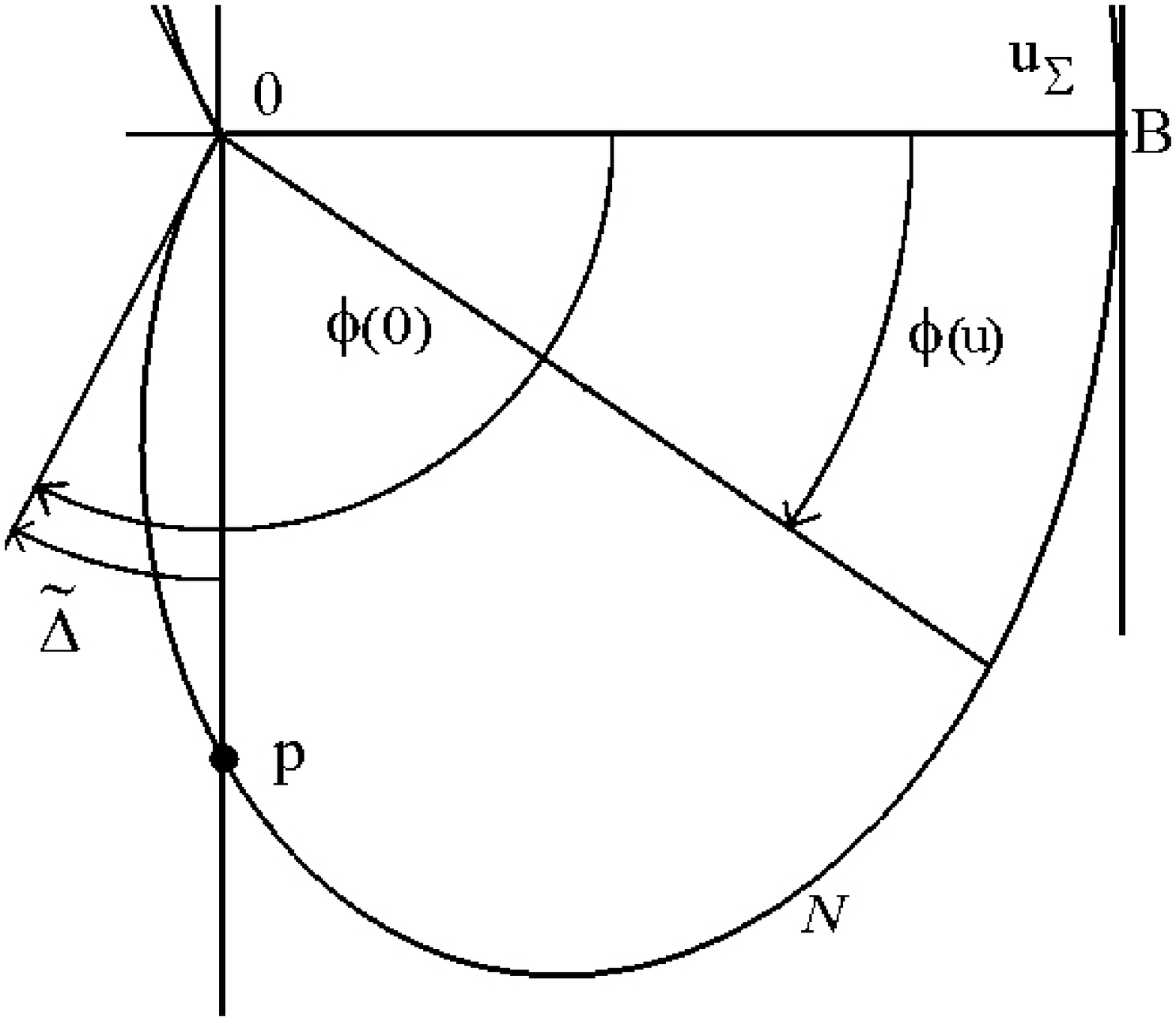,height=2in,width=2.5in,angle=0}
\caption{\label{trajectory} Part of the null geodesic \textit{N} given by (\ref{solution}) in the $u -  \phi$ subspace of (\ref{line}) with (\ref{factor}). The vertical axis is exaggerated for clarity. The line $OB$ is the line of apsides so the complete trajectory \textit{N} is obtained by adding the reflection of the part of \textit{N} shown about $OB$. $\tilde{\Delta}$ is explained in Table \ref{table1}. Exhibition of the entire orbit, without use of asymptotics, derives from the use of $u$ instead of \textsf{r}.}
\end{figure}

The non - radial null geodesic equation associated with (\ref{line}) given (\ref{factor}) can be written in the form
\begin{equation}
(\frac{d u}{d \phi})^2 =u_{_{\Sigma}}^2-2u_{_{\Sigma}}^3-u^2+2u^3
\label{u3}
\end{equation}
where $u \equiv m/\textsf{r}$ and the maximal value of $u$ is assumed fixed and given by $u_{_{\Sigma}} \equiv m/\textsf{r}_{_{\Sigma}} < 1/3$ irrespective of $\Lambda$ \cite{explain}.
Equation (\ref{u3}) is solved exactly, up to sign,  by \cite{exact}
\begin{equation}
\phi(u) = \sqrt{\frac{\Theta-k^2(u)}{-l(u)B}}\;\;\mathcal{F} \left(2\sqrt{\frac{u_{_{\Sigma}}-u}{A}},\sqrt{\frac{A}{B}}\right), \label{solution}
\end{equation}
where
\begin{equation}
\Theta \equiv (1-2u_{_{\Sigma}})(1+6u_{_{\Sigma}})>1,\label{theta}
\end{equation}
\begin{equation}
A \equiv 6u_{_{\Sigma}}-1+\sqrt{\Theta}>6u_{_{\Sigma}}>0,\label{A}
\end{equation}
\begin{equation}
B \equiv 6u_{_{\Sigma}}-1-\sqrt{\Theta}<2(3u_{_{\Sigma}}-1)<0,\label{B}
\end{equation}
\begin{equation}
k(u) \equiv 4u+2u_{_{\Sigma}}-1\label{k}
\end{equation}
so that
\begin{equation}
\sqrt{\Theta}+k(u)>2u_{_{\Sigma}}>0,\label{theta+k}
\end{equation}
and
\begin{equation}
\sqrt{\Theta}-k(u)>2(1-3 u_{_{\Sigma}})>0.\label{theta-k}
\end{equation}
The function $l(u)$ is given by \cite{llimits}
\begin{equation}
l(u) \equiv -2u_{_{\Sigma}}^2+u_{_{\Sigma}}-2u_{_{\Sigma}}u+u-2u^2>0\label{l}
\end{equation}
and $\mathcal{F}$ is the incomplete elliptic integral of the first kind of purely imaginary modulus \cite{abram}.
The solution (\ref{solution}) is shown in Figure \ref{trajectory}. Note that there is an apse at $B$ where $u=u_{_{\Sigma}}$ and $\phi(u_{_{\Sigma}})=0$  $\forall$  $u_{_{\Sigma}}<1/3$ \cite{cosmo}.

\bigskip

For $u=0$ we have
\begin{equation}
\phi(0) = 2\sqrt{\frac{2}{-B}}\;\; \mathcal{F} \left(2\sqrt{\frac{u_{_{\Sigma}}}{A}},\sqrt{\frac{A}{B}}\right). \label{solution0}
\end{equation}
Along the orbit, $u_{p}$ is distinguished by the condition
\begin{equation}
\frac{\pi}{2} = \sqrt{\frac{\Theta-k^2(u_{p})}{-l(u_{p})B}}\;\; \mathcal{F}\left(2\sqrt{\frac{u_{_{\Sigma}}-u_{p}}{A}},\sqrt{\frac{A}{B}}\right). \label{solutionp}
\end{equation}
Table \ref{table1} gives a numerical summary for grazing incidence with the Sun based on (accurate) numerical approximation to the exact solution (\ref{solution}).
\begin{table}
\caption{\label{table1}$u_{_{\Sigma}}=u_{_{\bigodot}},\;\;\;\;u$ variable \cite{number}.}
\begin{ruledtabular}
\begin{tabular}{ll}
  \textbf{Quantity (units)} & \textbf{Value $\simeq$}\\
  $\phi(0)$ (rad) & $1.57080\; rad$\\
  $2\phi(0)-\pi \equiv 2\tilde{\Delta}$ $('')$& $1''.75051$\\
  $4u_{_{\bigodot}}$ $('')$ & $2\tilde{\Delta} -0''.722465\; 10^{-5}$\\
  $4u_{_{\bigodot}}$+$u_{_{\bigodot}}^2(\frac{15 \pi}{4}-4)$ $('')$& $2\tilde{\Delta} -0''.336958\;10^{-10}$\\
  $\phi(u_{p})=\frac{\pi}{2}$, $p \equiv \frac{u_{_{\bigoplus}}}{u_{p}}$  & $p=1096.38$\\
  $\phi(u_{_{\bigoplus}})$ (rad) & $1.56615\; rad$ \\
  $2\phi(u_{_{\bigoplus}})-\pi$ $('')$& $-1917''.49$\\
\end{tabular}
\end{ruledtabular}
\end{table}

It is fair to say that the vast majority of texts (new and old) consider only equation (\ref{u3}), in the equivalent form
\begin{equation}\label{standard}
\frac{d^2 u}{d \phi^2}+u=3 u^2,
\end{equation}
by way of a first order approximation.
This procedure gives rise to the classic total deflection $4u_{_{\bigodot}}$ (for $\Lambda=0$) which, as Table \ref{table1} shows, is an excellent approximation to the total deflection defined by $2\tilde{\Delta}$. (Indeed, the second order approximation, given by $4u_{_{\bigodot}}$+$u_{_{\bigodot}}^2(\frac{15 \pi}{4}-4)$, is very much better still! See also Appendix A.) However, we do not live at $u=0$, or even at $u_{p}$ (in the inner Oort cloud) but at $u_{_{\bigoplus}}$ which has to be considered far away from $u=0$ in the sense that $\phi(0)-\phi(u_{_{\bigoplus}}) \simeq 959''.62$. Presentations that rely on considerations of (\ref{u3}) (or (\ref{standard})) alone are, therefore, misleading \cite{exception}. Yet, ``the bending of light" $\simeq 1''.75$ has been measured \cite{observation}, and without going to $u=0$ ! To understand this it is necessary to give an operational definition to ``the bending of light" which we now consider.

\section{The Bending of Light}
The ``bending of light" ($\psi$) is defined here to be the combination of the deflection prior to the apse plus the deflection after the apse,
\begin{equation}\label{deflection}
    \psi \equiv \Delta_{in}+\Delta_{out}.
\end{equation}

Prior to the apse we have
\begin{equation}\label{emitter}
    \Delta_{in} \equiv \phi(u_{e})-\frac{\pi}{2},
\end{equation}
where $e$ stands for the emitter and $\phi(u_{e})$ follows from (\ref{solution}). If $u_{e}=0$, and  for grazing incidence with the Sun, $\Delta_{in}=\tilde{\Delta}$. Approximations to $\Delta_{in}$ are carried out in Appendix A.

\bigskip

\begin{figure}[ht]
\epsfig{file=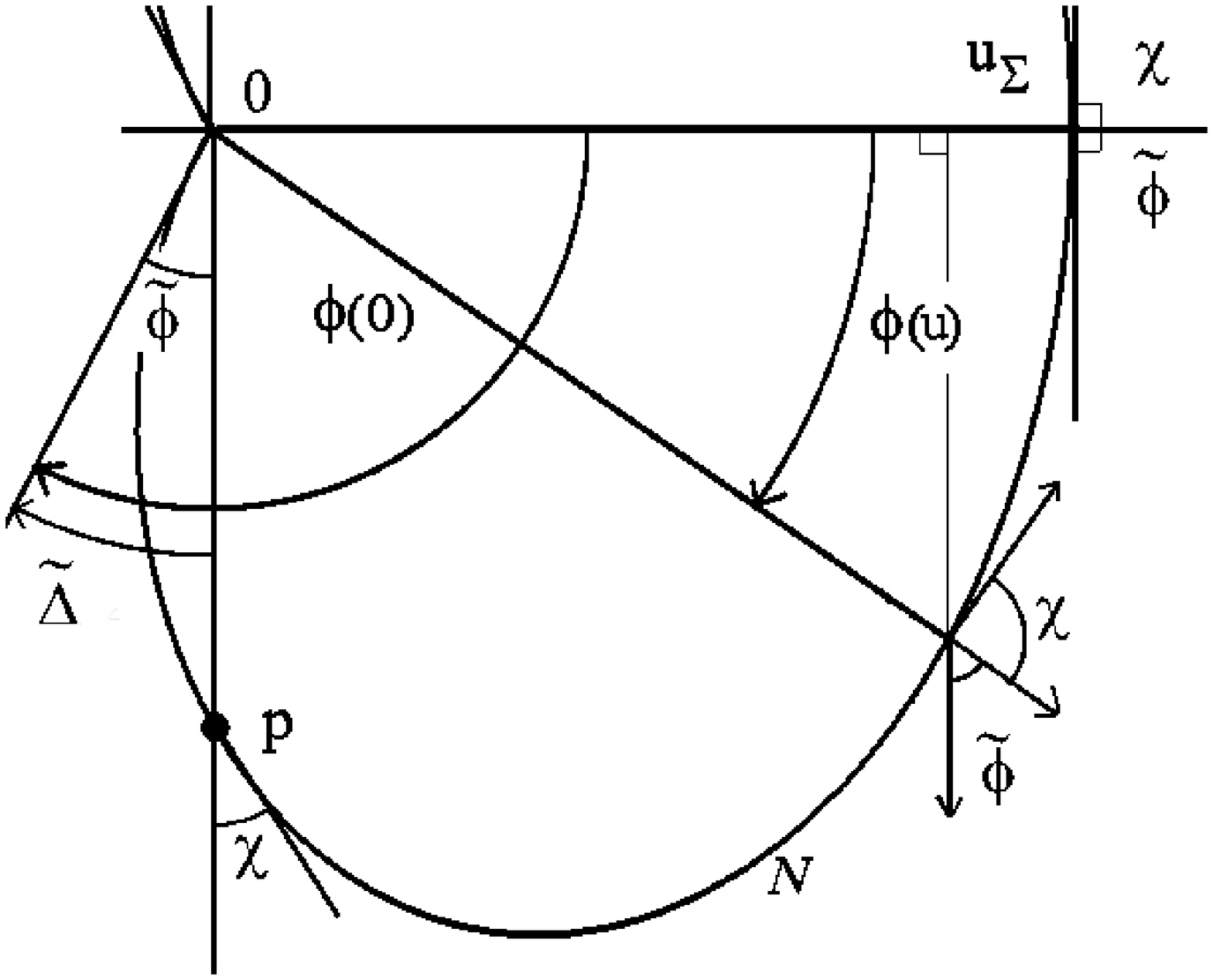,height=2.5in,width=3in,angle=0}
\caption{\label{evolution}Demonstration of the angles $\chi$ and $\tilde{\phi}$. The evolution of these angles for various observers along \textit{N} is also shown. The angle $\chi$ shrinks monotonically along \textit{N} from the apse, where $\chi=\pi/2$, to $O$ where $\chi=0$. The angle $\tilde{\phi}$ shrinks monotonically along \textit{N} from the apse, where $\tilde{\phi}=\pi/2$, to $p$ where $\tilde{\phi}=0$ and then increases to $\tilde{\Delta}$ at $O$. This is discussed further below.}
\end{figure}

\begin{figure}[ht]
\epsfig{file=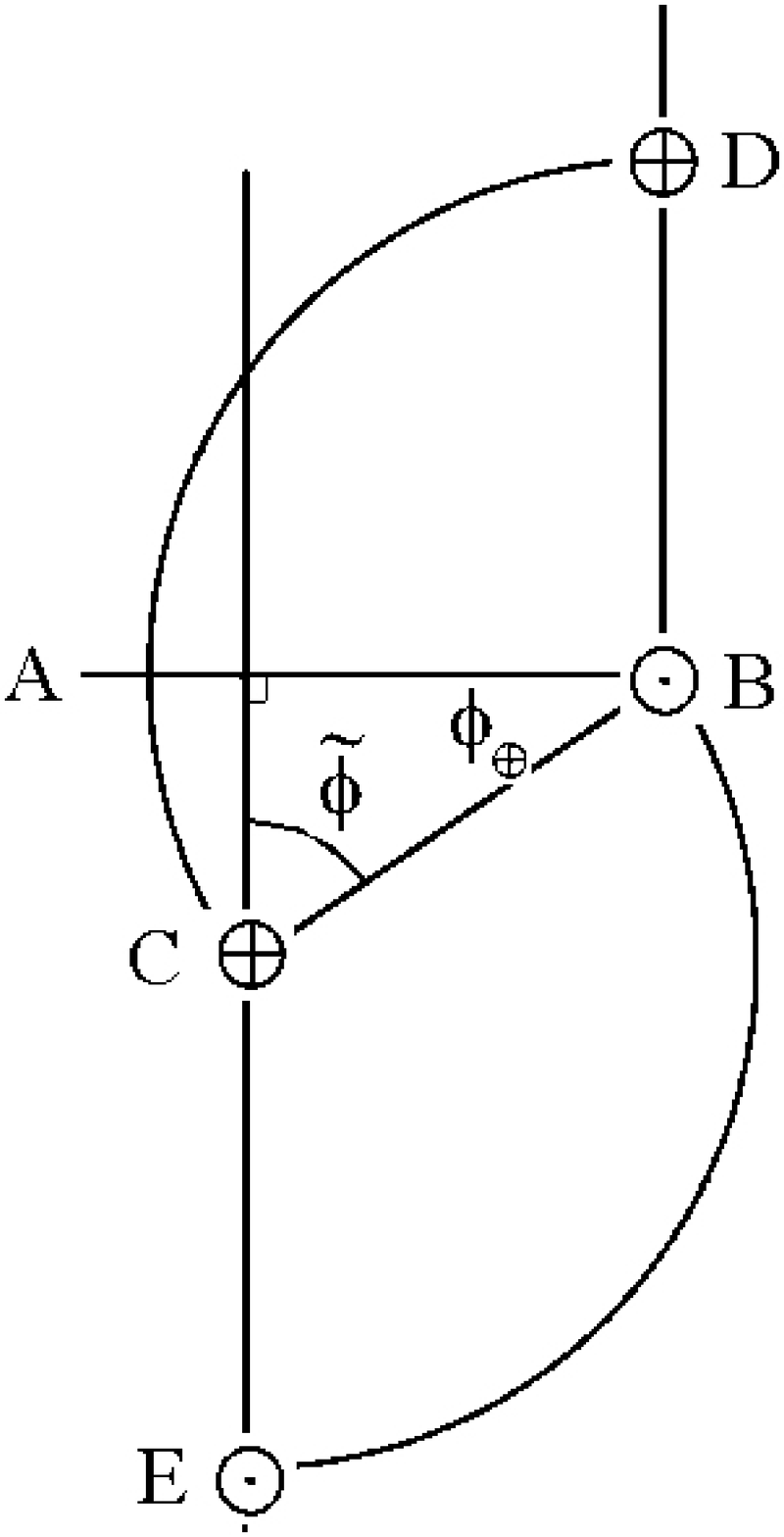,height=3in,width=2in,angle=0}
\caption{\label{phi}The undeflected position of a star is defined to be that position determined by radial null geodesics from the star. In the Figure $AB$ is the line of apsides. The perpendicular becomes a radial null geodesic on rotating the Sun through $\pi-\tilde{\phi}$ about the Earth, or equivalently, the Earth about the Sun by $\pi/2+\phi(u_{_{\bigoplus}})$.  We have assumed no parallax. The theoretical value for $\tilde{\phi}$ is then $\pi/2-\phi(u_{_{\bigoplus}})$. In practice, in the classical experiment, the undeflected positions are used to form a template and the time for alignment is $\sim 182.4$ days (let us say 6 months).}
\end{figure}

The deflection $\Delta_{out}$, unlike $\Delta_{in}$, involves the process of observation and the difference between two angles: $\chi$, the angle between the tangent to the null geodesic \textit{N} and the direction of the deflector as measured by a timelike observer on an orbit of constant $u$ in the $u-\phi$ subspace of (\ref{vacuum}) (the measured position of a star relative to the center of the Sun during a total solar eclipse in the classic bending of light experiment), and $\tilde{\phi}$, the angle defined between the position of the deflector and the perpendicular to the line of apsides. This relates the position of the star $\sim$ 6 months after the eclipse as explained below. The angles are shown in Figure \ref{evolution} along with their evolution along \textit{N} for various observers. The relationship of $\tilde{\phi}$ to theory and observation is explained in Figure \ref{phi}.

We have
\begin{equation}\label{Delta}
\Delta(u)_{out} \equiv \chi(u)-\tilde{\phi}(u)=\chi(u)+\phi(u)-\frac{\pi}{2},
\end{equation}
where $\phi(u)$ is given by (\ref{solution}). It is clear from Figure \ref{evolution} that $\Delta(u_{_{\Sigma}})_{out}=0$, $\Delta(u_{p})_{out}=\chi(u_{p})$ and $\Delta(0)_{out}=\phi(0)-\pi/2 \equiv \tilde{\Delta}$ (for a solar deflector). We are interested in $\Delta(u_{_{\bigoplus}})_{out}$.

\bigskip

\begin{figure}[ht]
\epsfig{file=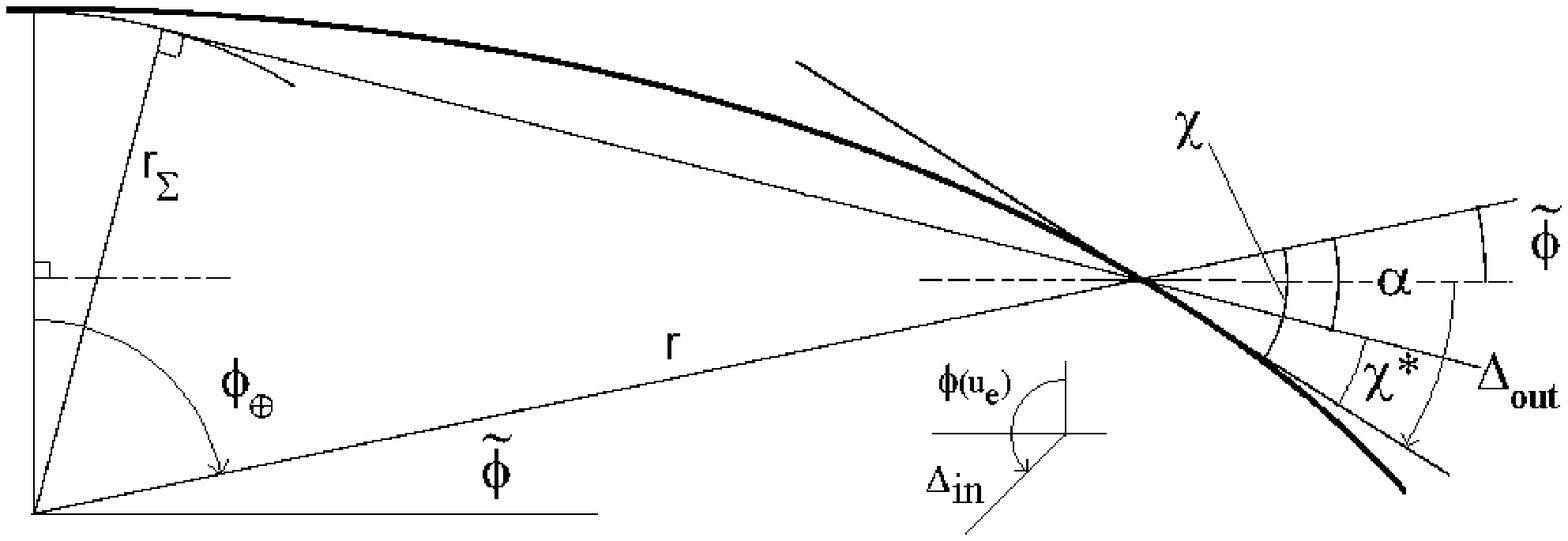,height=2.7in,width=6in,angle=0}
\caption{\label{chi}Deflection in the $\textsf{r} - \phi$ plane showing the variation of the reference point chosen for the definition of the measured angle $\chi$. As an example,  for $\chi$ the reference point is the center of the sun but for $\chi^{*}$ the reference point is the edge of the solar disc. Both $\Delta_{in}$ (shown in insert) and $\Delta_{out}$ simply measure the deviation from the horizontal (straight - line motion).}
\end{figure}

It is important to note that the reference point chosen for the definition of the measured angle $\chi$ does not in fact change $\Delta(u)_{out}$. For some other reference point we have
\begin{equation}\label{Deltamod}
\Delta(u)_{out} = \chi^{*}(u)+\alpha-\tilde{\phi}(u)=\chi^{*}(u)+\alpha+\phi(u)-\frac{\pi}{2},
\end{equation}
where the angles are shown in Figure \ref{chi}. The angle $\chi^{*}(u)$ is measured, $\phi(u)$ follows from (\ref{solution}) and $\alpha$ from the geometry as shown. In all cases $\chi=\chi^{*}+\alpha$. These angles are measured and calculated in the $\theta=\pi/2$, $\;t=const$ subspace of (\ref{line}) with (\ref{factor}) in which we use the standard formula
\begin{equation}\label{angle}
    \cos(\iota)=\frac{d_{i}\delta^{i}}{\sqrt{d_id^i}\sqrt{\delta_i\delta^i}}
\end{equation}
for the angle $\iota$ between the two directions defined by $d^{i}$ and $\delta^{i}$.

\bigskip

From (\ref{line}) with (\ref{factor}) and (\ref{angle}) we find the theoretical prediction for $\chi$ \cite{kn},
\begin{equation}\label{chicalc}
\sin(\chi)=\left(\frac{u}{u_{_{\Sigma}}}\sqrt{\frac{f}{f_{_{\Sigma}}}}\right)
\end{equation}
where we have written $\chi$ for $\chi(u)$, $\;f$ for $f(u)$ and $f_{_{\Sigma}}$ for $f(u_{_{\Sigma}})$.
(Because we have chosen a timelike observer on an orbit of constant $u$ to measure $\chi$, we have the restriction $u \geq u_{\mathcal{H}}$. This does not change in any significant way the current presentation.) Approximations to $\chi$ are carried out in Appendix B. Now if, as shown in Figure \ref{chi}, we were to choose $\chi^{*}$, a similar calculation gives
\begin{equation}\label{chistar}
\cos(\chi^{*})=\frac{u_{_{\Sigma}}}{u}\left(\frac{\sqrt{(1-(\frac{u}{u_{_{\Sigma}}})^2)(f_{_{\Sigma}}-(\frac{u}{u_{_{\Sigma}}})^2f)}+(\frac{u}{u_{_{\Sigma}}})^2f}
{\sqrt{f_{_{\Sigma}}}\sqrt{(\frac{u}{u_{_{\Sigma}}})^2+f-1}} \right)
\end{equation}
with
\begin{equation}\label{alpha}
\cos(\alpha)=\frac{\sqrt{(\frac{u}{u_{_{\Sigma}}})^2-1}}{\sqrt{(\frac{u}{u_{_{\Sigma}}})^2+f-1}}.
\end{equation}
Form (\ref{chicalc}), (\ref{chistar}), and (\ref{alpha}) it can be shown that $\chi=\chi^{*}+\alpha$. Now we could choose the reference direction to be the measured position of the deflected star in which case $\chi^{*}=0$ and $\alpha$, in numerical value, $=\chi$. With the foregoing understood, we emphasize the choice $\chi$ over $\chi^{*}$ just for simplicity in what follows. First, however, it is instructive to look at the weak - field limit $f \rightarrow 1$. In this limit we obtain $\chi=\arcsin(\textsf{r}_{\Sigma}/\textsf{r})$, $\chi^{*}=0$ and $\alpha=\arcsin(\textsf{r}_{\Sigma}/\textsf{r})$ as we would expect.

\bigskip

From (\ref{solution}),  (\ref{Delta}) and (\ref{chicalc}), with all currently acceptable values of $\Lambda$, and for grazing incidence with the Sun, we find \cite{onehalf}
\begin{equation}\label{finalearth}
\Delta(u_{_{\bigoplus}})_{out} \simeq 1''.75049\;\; (\frac{1}{2}).
\end{equation}
This result is best understood in the following way:
\begin{equation}\label{phiearth}
-\tilde{\phi}(u_{_{\bigoplus}})=\phi(u_{_{\bigoplus}})-\frac{\pi}{2} \simeq -1917''.49\;\; (\frac{1}{2})
\end{equation}
and
\begin{equation}\label{chiearth}
\chi(u_{_{\bigoplus}}) \simeq +1919''.24\;\; (\frac{1}{2}).
\end{equation}
It is, of course, crucial here that we evaluate $\phi(u_{_{\bigoplus}})$ from (\ref{solution}). The almost exact cancelation by addition of these two contributions at $u_{_{\bigoplus}}$, along with the assumption that $u_{e}=0$, gives the famous result $\psi(u_{_{\bigoplus}}) \simeq 1''.75$. This cancelation is, of course, not unique to the Earth \cite{serendipity}. Table \ref{table2} demonstrates this.
\begin{table}
\caption{\label{table2}$u_{_{\Sigma}}=u_{_{\bigodot}},\;\;\;\;u$ variable \cite{numbernew}.}
\begin{ruledtabular}
\begin{tabular}{llll}
  \textbf{Distance (AU)} & \textbf{$2\phi-\pi$} & \textbf{$2\chi$} & \textbf{$2\Delta_{out}$} \\
  1/214.75 &                   -630397      &            630398      &          0.07467\\
  1/214  &                 -609289         &         609290      &          0.16402\\
  1/213    &                  -592452       &            592452      &           0.23500\\
  1/212    &                   -579625        &          579626         &       0.28881 \\
  1/210    &                  -559334         &         559335            &    0.37335\\
  1/100    &                   -199621          &        199623            &    1.54953\\
  1/10     &                   -19197.5        &         19199.2         &      1.74862\\
  Mercury \cite{semi} &    -4956.37           &      4958.12          &     1.75038\\
  Venus       &                  -2651.59     &            2653.34      &         1.75048\\
  Earth      &                   -1917.49       &          1919.24       &        1.75049\\
  Mars          &                -1257.87          &       1259.62         &      1.75050\\
  Jupiter      &                 -367.030      &           368.781           &    1.75051\\
  Saturn          &              -198.545        &         200.295           &    1.75051\\
  Uranus          &              -98.0566        &         99.8071           &    1.75051\\
  Neptune       &                -62.0036        &         63.7541          &     1.75051\\
  50       &                   -36.6341       &          38.3846           &    1.75051\\
  500      &                   -2.08795         &         3.83846           &     1.75051\\
  1096.38         &            0               &        1.75051        &       1.75051\\
  5000     &                   1.36666          &         0.383846      &         1.75051
\end{tabular}
\end{ruledtabular}
\end{table}

\bigskip

The vast majority of texts (new and old) consider only the deflection of a single source and argue that the observed bending of light derives from the fact that we can set both the emitter and observer at $u=0$ and so
\begin{equation}\label{standardre}
\psi=2\Delta_{in}=2\phi(u \simeq 0)-\pi \equiv 2\tilde{\Delta} \simeq 1''.75.
\end{equation}
Whereas this argument arrives at a good answer, on the basis of the exact solution to (\ref{u3}) I have shown that in fact (\ref{phiearth}) holds
and so these approximate arguments leading to (\ref{standardre}) do not adequately explain the measured deflection of light at the Earth when a single source is observed. If we set $u_{e}=0$, so that $\Delta_{in}=\tilde{\Delta}$, we must still account for $\Delta_{out}$ and for that I have shown that the measured value is the result of the almost perfect cancelation of two terms, (\ref{phiearth}) and (\ref{chiearth}), one of which is seldom considered, namely (\ref{chiearth}). It is also this term that formally introduces $\Lambda$ by way of (\ref{chicalc}) and (\ref{factor}). This cancelation is discussed in Appendix C. If one chooses not to use the center of the sun as the reference point for the measurement of $\chi$, this cancelation remains. It is then the cancelation of $\chi^{*}+\alpha$ with $\tilde{\phi}$ that explains measured deflection of light.

\bigskip

\section{Wide-angle deflections}
Nowadays the classical bending of light test is extended over the entire sky and is used with interferometric methods at radio wavelengths to put (remarkable) limits on the PPN parameter $\gamma$ \cite{shapiro}.  This is done on the basis of the Shapiro - Ward formula \cite {shapward}, which in our notation reads as
\begin{equation}\label{shward}
    \Psi \simeq 2 u_{_{\Sigma}}(1+\cos (\chi(u))).
\end{equation}
Note, however, that in (\ref{shward}) $\chi(u)$ is often taken as an observed quantity, and not derived from (\ref{chicalc}).
Here we compare this relation with $\psi$ as defined in (\ref{deflection}). The situation is summarized in Table \ref{table3} where larger values of the apsidal distance are considered up to $\chi=\pi/2$ and the observer is at $u=u_{_{\bigoplus}}$. Whereas there is no measurable difference in the two definitions ($\psi$ and $\Psi$), they are not equivalent and the cancelation discussed in the previous section plays the dominant role. This is also discussed in Appendix C.

\bigskip

\section{Differential Measurements}
\begin{figure}[ht]
\epsfig{file=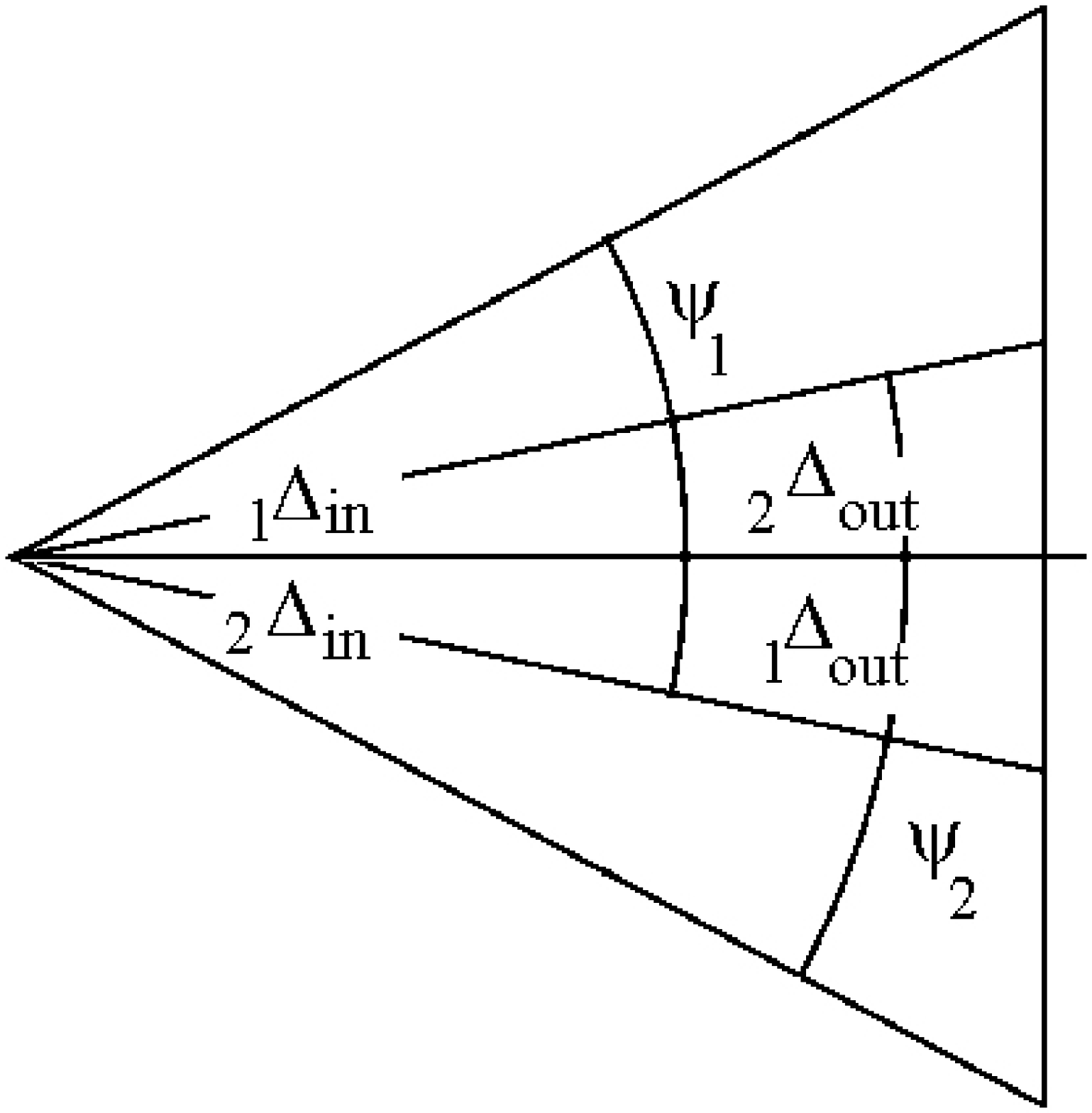,height=2in,width=2.5in,angle=0}
\caption{\label{differentialf}Constructing the differential measurement $\delta$.}
\end{figure}

In practice it is the measurement of differential deflections in opposition that is also of importance \cite{plebanski}. Consider two sources, 1 and 2, which have associated with them the same $u_{_{\Sigma}}$ (but not necessarily for grazing incidence with the Sun) that are seen on opposite sides of the deflector (in the classic case, during a total eclipse). Then the differential deflection is given by
\begin{equation}\label{differential}
\delta \equiv \psi_{1}-\psi_{2}
\end{equation}
as shown in Figure \ref{differentialf}.
Now by construction
\begin{equation}\label{out}
    _{1}\Delta_{out}=\;_{2}\Delta_{out}
\end{equation}
and (in view of the evolution of the null geodesics about the deflector)
\begin{equation}\label{in}
     _{1}\Delta_{in}=-_{2}\Delta_{in}
\end{equation}
so that
\begin{equation}\label{delta}
    \delta=2\;_{1}\Delta_{in}.
\end{equation}
For an emitter at $u_{e}=0$ we have the rigorous result
\begin{equation}\label{rig}
    \delta=2\phi(0)-\pi.
\end{equation}
Of particular note is the fact that for this configuration since $\Delta_{out}$ cancels out,
$\Lambda$ plays no role, and further, the usual approximate argument gives the rigorously correct answer up to numerical evaluation.

\bigskip

More generally, when the associated $u_{_{\Sigma}}$ are not equal, but we continue to consider opposed sources and set $u_{e}=0$,
\begin{equation}\label{general}
    \delta=(\phi_{1}(0)+\phi_{2}(0)-\pi)+(\chi_{1}(u_{_{\bigoplus}})-\chi_{2}(u_{_{\bigoplus}}))
\end{equation}
where the $\phi(0)$ are given by (\ref{solution0}) and the $\chi$ by (\ref{chicalc}). Whereas $\Lambda$ does not enter the first term, it does not cancel from the second except in the limit $_{1}u_{\Sigma}=\;_{2}u_{\Sigma}$ which reproduces (\ref{rig}).

\bigskip

When the associated $u_{_{\Sigma}}$ are not equal, and the sources are not opposed, the differential deflection becomes
\begin{equation}\label{general1}
    \delta=(\phi_{1}(0)-\phi_{2}(0))+(\chi_{1}(u_{_{\bigoplus}})-\chi_{2}(u_{_{\bigoplus}}))
\end{equation}
where again the $\phi(0)$ are given by (\ref{solution0}) and the $\chi$ by (\ref{chicalc}) and we have taken $u_{e}=0$. Whereas $\Lambda$ does not enter the first term, it does not cancel from the second except in the limit $_{1}u_{\Sigma}=\;_{2}u_{\Sigma}$ which gives $\delta=0$ (we have one source).

\bigskip

\section{Discussion}
The standard argument for explaining the measured value of the deflection of light is to set the emitter and observer at $u=0$ so that the total deflection for a single source for grazing incidence with the Sun is given by $(2\phi(0)-\pi)|_{u_{_{\bigodot}}}$, which, when approximated by first order corrections to the null geodesic equation, gives $4u_{_{\bigodot}}$ $('')$. For wide-angle deflections a similar argument leads to (\ref{shward}). On the basis of the mathematically exact solution to the classical bending problem I have shown that these are excellent approximations. However, as regards \textit{explaining} the measured deflection of light, these approximate arguments fall short in the sense that the bending prior to and after an apse are fundamentally different. The post apsidal bending involves the process of measurement and is dominated by the cancelation of two terms, one of which is not considered in the usual approximate argument. Whereas these more involved arguments may not be numerically important, conceptually they are.

\bigskip

Aside form these details, the present discussion differs from others in the sense that analytic approximations to the exact solution (\ref{solution}) is not central to the discussion, as is the usual case, but rather (\ref{solution}) is viewed as an elementary function in the sense that it can be trivially approximated numerically to arbitrary accuracy using modern computational platforms. However, approximations have been included, relegated to the Appendices, as these serve to amplify and ``explain" the results obtained.

\begin{acknowledgments}
This work was supported by a grant from the Natural Sciences and
Engineering Research Council of Canada.
\end{acknowledgments}

\begin{widetext}

\begin{table}
\caption{\label{table3}$u_{_{\Sigma}}$ variable,$\;\;\;\;u=u_{_{\bigoplus}}$ .}
\begin{ruledtabular}
\begin{tabular}{lllllllll}
  \textbf{$\textsf{r}_{\Sigma}/R_{_{\bigodot}}$}& $\chi (^{0})\sim$& \textbf{$\phi-\pi/2\; ('')$} & \textbf{$\chi\; ('')$} & \textbf{$\Delta_{out}\; ('')$}&\textbf{$\Delta_{in}\; ('')$}&\textbf{$\psi\; ('')$}&\textbf{$\Psi\; ('')$}\\
  1    &    0.27   &  -958.744 &  959.619 & 0.875246 & 0.875255  &   1.75050  &    1.75049\\
  1.5   &   0.40  &   -1438.85 &  1439.43 & 0.583489 & 0.583503   &  1.16699   &   1.16699\\
  1.75  &   0.47  &   -1678.84 &  1679.34 & 0.500128 & 0.500145   &  1.00027   &   1.00027\\
  2    &    0.53   &  -1918.82 &  1919.26 & 0.437608 & 0.437627   &  0.875235  &   0.875233\\
  2.5  &    0.67  &   -2398.74 &  2399.09 & 0.350078 & 0.350101  &   0.700179  &   0.700178\\
  3     &   0.80   &  -2878.65 &  2878.94 & 0.291723 & 0.291751  &   0.583474  &   0.583474\\
  3.5   &   0.93  &   -3358.55 &  3358.80 & 0.250039 & 0.250072  &   0.500111  &   0.500111\\
  4    &    1.07  &   -3838.46 &  3838.68 & 0.218775 & 0.218813  &   0.437588  &   0.437588\\
  4.5   &   1.20  &   -4318.39 &  4318.58 & 0.194458 & 0.194501  &   0.388959  &   0.388958\\
  5     &   1.33  &   -4798.33 &  4798.50 & 0.175003 & 0.175050  &   0.350054  &   0.350053\\
  5.5   &   1.47  &   -5278.29 &  5278.45 & 0.159085 & 0.159137 &    0.318222  &   0.318221\\
  6    &    1.60  &   -5758.29 &  5758.43 & 0.145819 & 0.145875 &    0.291694  &   0.291694\\
  6.5   &   1.73 &    -6238.31 &  6238.44 & 0.134593 & 0.134593  &   0.269247  &   0.269247\\
  7    &    1.87  &   -6718.36 &  6718.49 & 0.124970 & 0.125036  &   0.250006  &   0.250006\\
  8    &    2.13  &   -7678.58 &  7678.69 & 0.109331 & 0.109407  &   0.218737  &   0.218737\\
  9    &    2.40  &   -8638.95 &  8639.05 & 0.0971650 &0.0972503 &   0.194415  &   0.194415\\
  10   &    2.67  &   -9599.52 &  9599.60 & 0.0874304 &0.0875252 &   0.174956  &   0.174956\\
  15   &    4.00  &   -14405.9 &  14405.9 & 0.0582079 &0.0583501 &   0.116558  &   0.116558\\
  20   &    5.34  &   -19220.0  & 19220.1 & 0.0435727 &0.0437626  &  0.0873353 &   0.0873353\\
  30   &    8.02  &   -28882.7 &  28882.7 & 0.0288895 &0.0291751  &  0.0580646 &   0.0580646\\
  40   &    10.7  &   -38609.6 &  38609.6 & 0.0214991 &0.0218813 &   0.0433804 &   0.0433804\\
  50   &    13.5  &   -48424.3 &  48424.3 & 0.0170248 &0.0175050  &  0.0345299 &   0.0345299\\
  100   &   27.7  &   -99811.3 &  99811.3 & 0.00774762 &0.00875252 & 0.0165001  &  0.0165001\\
  125   &   35.6  &   -128013  &  128013  & 0.00569625 &0.00700201 & 0.0126983  &  0.0126983\\
  150  &    44.3  &   -159317  &  159317 &  0.00417929 &0.00583501 & 0.0100143  &  0.0100143\\
  200  &    68.5  &   -246628  &  246628  & 0.00160335 &0.00437626 & 0.00597961  & 0.00597961\\
  210  &    77.7 &    -279667  &  279667 &  0.000888921 &0.00416787 &0.00505679  & 0.00505679\\
 214.95 &  90   &    -323867  &  323867  & 0.000002635 &0.00407197 &0.00407460  & 0.00407460
\end{tabular}
\end{ruledtabular}
\end{table}

\newpage

\appendix

\section{Approximations to $\Delta_{in}$}
If
\begin{eqnarray}\label{defdeltainu}
   u_{e} \ll u_{_{\Sigma}} \nonumber
\end{eqnarray}
then from (\ref{solution}) it follows that
\begin{eqnarray}\label{defdeltainphi}
   \Delta_{in} \simeq \phi(0)-\frac{\pi}{2}. \nonumber
\end{eqnarray}
If $\textsf{r}_{_{\Sigma}} \gg m$ we can take a Taylor series of $\phi(0)-\pi/2$ about $u_{_{\Sigma}}=0$ which gives

\begin{eqnarray}\label{taylor}
\phi(0)-\pi/2 \simeq 2\,u_{_{\Sigma}}+ \left(  -2 + {\frac {15}{8}}\,\pi\right) {u_{_{\Sigma}}}^{2}+ \left( {\frac {
61}{3}}-{\frac {15}{4}}\,\pi  \right) {u_{_{\Sigma}}}^{3}+ \left( -65+{\frac {3465
}{128}}\,\pi  \right) {u_{_{\Sigma}}}^{4}+ \left( {\frac {7783}{20}}-{\frac {3465}
{32}}\,\pi  \right) {u_{_{\Sigma}}}^{5}+\\ \nonumber \left( -{\frac {21397}{12}}+{\frac {
310695}{512}}\,\pi  \right) {u_{_{\Sigma}}}^{6}+ \left( {\frac {544045}{56}}-{
\frac {765765}{256}}\,\pi  \right) {u_{_{\Sigma}}}^{7}+ \left( -{\frac {400353}{8}
}+{\frac {530675145}{32768}}\,\pi  \right) {u_{_{\Sigma}}}^{8}+ \\ \nonumber \left( {\frac {
1094345069}{4032}}-{\frac {350975625}{4096}}\,\pi  \right) {u_{_{\Sigma}}}^{9}+
 \left( { -{\frac {3274477761}{2240}+\frac {61238992815}{131072}}\,\pi}
 \right) {u_{_{\Sigma}}}^{10}
 +\mathcal{O}(u_{_{\Sigma}}^{11}),
\end{eqnarray}
and it is a trivial matter to go to higher orders. The first six terms in (\ref{taylor}) have been given previously \cite{keeton}. In many applications the first term alone represents an adequate approximation.
\section{Approximations to $\chi$}
From (\ref{factor}) and (\ref{chicalc}) it follows that
\begin{equation}\label{newchi}
 \cos(\chi(u))^2=\frac{u_{_{\Sigma}}^2-2u_{_{\Sigma}}^3-u^2+2u^3}{u_{_{\Sigma}}^2-2u_{_{\Sigma}}^3-\lambda}
\end{equation}
where
\begin{eqnarray}\label{newlambda}
    \lambda \equiv \frac{\Lambda m^2}{3}
\end{eqnarray}
and $u$ corresponds to the observer.
Equation (\ref{newchi}) is exact. Let us suppose that $\textsf{r}_{_{\Sigma}}\gg 2m$ so that
\begin{eqnarray}\label{sigmalimit}
    u_{_{\Sigma}}^2-2u_{_{\Sigma}}^3 \simeq u_{_{\Sigma}}^2,
\end{eqnarray}
and further, let us suppose that for the observer $\textsf{r} \gg 2m$, so that
\begin{eqnarray}\label{ulimit}
    -u^2+2u^3 \simeq -u^2.
\end{eqnarray}
Then (\ref{newchi}) takes the form
\begin{equation}\label{newchilim}
 \cos(\chi(u))^2 \simeq \frac{u_{_{\Sigma}}^2-u^2}{u_{_{\Sigma}}^2-\lambda}.
\end{equation}
If $\lambda=0$ we arrive at the obvious result
\begin{equation}\label{newchilim0}
\sin(\chi(u)) \simeq \frac{u}{u_{_{\Sigma}}}.
\end{equation}
If $\lambda>0$ let us assume that
\begin{equation}\label{lamdalimit}
    9m^2\Lambda \ll 1,
\end{equation}
that is, $27 \lambda \ll 1$. Then, for the cosmological horizon $\mathcal{H}$, we obtain \cite{jhm}
\begin{equation}\label{cosmo}
    u_{\mathcal{H}}^2 \simeq \lambda
\end{equation}
so that we can rewrite (\ref{newchilim}) in the form
\begin{equation}\label{newchilimlambda}
     \cos(\chi(u))^2 \simeq \frac{u_{_{\Sigma}}^2-u^2}{u_{_{\Sigma}}^2-u_{\mathcal{H}}^2}.
\end{equation}
Under the assumption $\textsf{r}_{_{\Sigma}}\ll r_{\mathcal{H}}$, that is
\begin{equation}\label{horizon}
    u_{_{\Sigma}}^2 \gg u_{\mathcal{H}}^2,
\end{equation}
we again obtain the now somewhat less obvious result (\ref{newchilim0}).

\bigskip

To gauge the effect of $\Lambda$ consider
\begin{equation}\label{Psi}
    \psi_{_{\Lambda=0}} - \psi_{_{\Lambda \neq 0}},
\end{equation}
which reduces to
\begin{equation}\label{chidiff}
  \chi_{_{\Lambda=0}} - \chi_{_{\Lambda \neq 0}}.
\end{equation}
Putting this into a crude cosmological context, let us take the following fiducial values: $\Lambda \sim \alpha \;10^{-56}\; cm^{-2}$, $m \sim \frac{3}{2} \;\beta \;10^{18}\; cm$, $\beta$ in units of $10^{13} \;M_{_{\bigodot}}$, $u_{_{\Sigma}} \sim \frac{\beta}{2 \gamma} \;10^{-5}$, $\gamma$ in units of 100 Kpc, and $u \sim \frac{\beta}{2 \delta}\; 10^{-9}$, $\delta$ in units of Gpc. Here $\alpha, \beta, \gamma$ and $\delta$ are taken to be of order unity. Now (\ref{sigmalimit}) requires $\frac{\beta}{\gamma} \ll 10^{5}$, and (\ref{ulimit}) requires $\frac{\beta}{\delta} \ll 10^{9}$. For (\ref{lamdalimit}) we require $2 \alpha \beta^2 \ll 10^{19}$ and for (\ref{horizon}) $8 \alpha \gamma^2 \ll 27 \;10^{9}$. Relation (\ref{newchilimlambda}) should therefore be a reasonable approximation in this context. Note that we can not set $u \gg u_{\mathcal{H}}$ as this would require $8 \alpha \delta^2 \ll 270$. Gathering this all together, we have the following approximation,
\begin{equation}\label{approxlambda}
   \chi_{_{\Lambda=0}} - \chi_{_{\Lambda \neq 0}} \simeq \frac{1}{2}\; \frac{\sqrt{u_{_{\Sigma}}^2-u^2}}{u u_{_{\Sigma}}^2} \lambda + \mathcal{O}(\lambda^2) \simeq  \frac{1}{2}  \; \frac{\lambda}{u u_{_{\Sigma}}} +   \mathcal{O}(\lambda^2),
\end{equation}
where the last term assumes $\frac{\gamma}{\delta} \ll 10^4$. For unit values of $\alpha, \gamma$ and $\delta$ then $\chi_{_{\Lambda=0}} - \chi_{_{\Lambda \neq 0}} \sim 0^{''}.3$ which, in principle, is a measurable difference. However, since we can not set $u \gg u_{\mathcal{H}}$, the cosmological context has to be refined \cite{newlensing}.

\section{Approximations to $\Delta_{out}$}

If we start with the assumption $\textsf{r}_{_{\Sigma}} \ll \textsf{r}$, where $\textsf{r}$ corresponds to the observer, then
\begin{equation}\label{bogus}
u_{_{\Sigma}} \gg u
\end{equation}
and we again obtain $\phi(u) \simeq \phi(0)$ and also $\chi \simeq 0$ so that
\begin{equation}\label{outbad}
   \psi \simeq  2\phi(0)-\pi,
\end{equation}
as one finds in most texts. However, the approximation (\ref{bogus}) is inadequate. Rather, with
\begin{equation}\label{reallim}
  u_{_{\Sigma}} \geq u
\end{equation}
and $ u_{_{\Sigma}} \leq \sim 10^{-4}$ to the 6 figure accuracy reported in this paper we find
\begin{equation}\label{true}
   \phi(u)-\frac{\pi}{2}+\chi(u) \simeq 2\sqrt{u_{_{\Sigma}}^2-u^2}.
\end{equation}
This ``explains" the almost perfect cancelation of the two terms on the left and the remarkable accuracy of the Shapiro - Ward formula (\ref{shward}). It is only in the limit $u=0$ that one recovers (\ref{outbad}).
\end{widetext}


\begin{thebibliography}{}\label{sec:TeXbooks}
\bibitem[*]{email}{Electronic Address: lake@astro.queensu.ca}
\bibitem{rindler} W. Rindler and M. Ishak, Phys. Rev. D \textbf{76}, 043006 (2007).
\bibitem{lake}K. Lake, Phys. Rev. D \textbf{65}, 087301 (2002).
\bibitem{explain}The constancy of $u_{\Sigma}$ (as regards $\Lambda$), for grazing incidence, is set by junction conditions on the deflector as explained in \cite{lake}. Since we have assumed that $u \leq u_{_{\Sigma}}$, the form (\ref{solution}) extends up to the condition that the null geodesic is observed at the apse. This is considered further below.
\bibitem{exact} The fact that (\ref{u3}) can be solved exactly in terms of Elliptic integrals is well known. See, for example, S. Chandrasekhar,\textit{ The Mathematical Theory of Black Holes}, (Oxford Universirty Press, Oxford 1983). Whereas many works are primarily concerned with analytical approximations to (\ref{solution}), these are not of central concern here. See also G. V. Kraniotis, Class. Quantum Grav. \textbf{22}, 4391 (2005).
\bibitem{llimits}For $1/4 \leq u_{_{\Sigma}}<1/3$,  $l(u)>2 u_{_{\Sigma}}(1-3 u_{_{\Sigma}})>0$ and for $u_{_{\Sigma}} \leq 1/4$,  $l(u)> u_{_{\Sigma}}(1- 2 u_{_{\Sigma}})>0$.
\bibitem{abram}See, for example, \textit{Handbook of Mathematical Functions}, Edited by M. Abramowitz and I. A. Stegun (Dover, New York, 1972) Section 17.4.17.
\bibitem{cosmo}The limit $u=0$ plays a central role in most arguments and this limit deserves special consideration for $\Lambda>0$. If the emitter lies within the cosmological horizon we have the restriction $u \geq u_{\mathcal{H}}$ where $u_{\mathcal{H}} \sim \sqrt{\Lambda m^2/3}$ for $\Lambda>0$ \cite{jhm}. Using current limits on $\Lambda$, for a solar mass deflector, we have $u_{\mathcal{H}} < 2.2\; 10^{-26}$. This does not change in any significant way any argument in this paper (but see Appendix B).
\bibitem{jhm} See, for example, M. J. Jaklitsch, C.Hellaby and D. R. Matravers, GRG \textbf{21} 941 (1989) (Appendix I).
\bibitem{number}We use the following \textit{adopted} values: $m_{_{\bigodot}} =1.47664\; 10^5 cm$, $\textsf{R}_{_{\bigodot}} = 6.9598\; 10^{10}\; cm$ and $AU=1.4959787\; 10^{13} cm$. I report only 6 digit accuracy,
    whereas the calculations were done with 40 digit accuracy under Maple, assuming, of course, that the adopted values are defined to be exact to this degree. When viewing the Tables, one should be mindful of roundoff. In the real world, light deflection to 6 digit accuracy is of course influenced by other factors not considered here.
\bibitem{exception}Let us note here that a few presentations do go beyond a consideration of (\ref{standard}) alone. These include C. M. Will, \textit{Theory and experiment in gravitational physics}, (Cambridge University Press, Cambridge, 1993) and C. W. Misner, K. S. Thorne and J. A, Wheeler, \textit{Gravitation}, (Freeman, San Francisco, 1973).
\bibitem{observation} See C. M. Will, ``The Confrontation between General Relativity and Experiment",
Living Rev. Relativity \textbf{9},  (2006),  3. URL (cited on October 22, 2007):
\texttt{http://www.livingreviews.org/lrr-2006-3}.
\bibitem{kn}The text by B. O'Neill, \textit{Semi-Riemannian Geometry} (Academic Press, New York, 1983) gives (\ref{chicalc}) for general $f$ given by (\ref{factor}). The text by M. Kriele, \textit{Spacetime} (Springer, Berlin, 2001) gives the result for $\Lambda=0$. Curiously, in neither case is $\chi$ related to $\Delta$ which is calculated in these texts in the usual approximate way, $\Delta \approx \tilde{\Delta}$.
\bibitem{onehalf}The notation comes from the fact that the factor $\frac{1}{2}$ is exact.
\bibitem{numbernew} Note that $2 \Delta_{out}$ increases monotonically with distance even though this is not clear in the Table because of \cite{number}. See also Appendix C.
\bibitem{semi} For the planets I have used the semi-major axis.
\bibitem{serendipity} It is, however, dificult to ignore the wonderful serendipity that Table \ref{table2} provides. For the Earth, $2\chi$ is the date of the famous eclipse. Further, for the Earth, $2\phi-\pi$ is the date before this eclipse when Astronomer Royal Sir F. W. Dyson first pointed out the opportunity afforded by this eclipse for verifying Einstein's theory of gravitation (see F. W. Dyson, MNRAS \textbf{77}, 445 (1917)).
\bibitem{shapiro}See, S. S. Shapiro, J. L. Davis, D. E. Lebach and J. S. Gregory, Phys. Rev Lett. \textbf{92}, 121101 (2004).
\bibitem{shapward}See I. I. Shapiro, Science \textbf{157}, 806 (1967) and W. R. Ward, Ap. J. \textbf{162}, 345 (1970).
\bibitem{plebanski} This is sometimes referred to as ``Eddington's method". See, for example, J. Pleba\'{n}ski and A. Krasi\'{n}ski, \textit{General Relativity and Cosmology}, (Cambridge University Press, Cambridge, 2006). Since we have restricted our analysis to $\theta=\pi/2$, when we consider more than one source, the sources must be aligned radially through the centre of the deflector.
\bibitem{keeton} C. R. Keeton and A. O. Petters, Phys. Rev D \textbf{72 }, 104006 (2005) {\tt arXiv:gr-qc/0511019v1}
\bibitem{newlensing} For recent discussions of lensing with $\Lambda$ see M. Ishak, W. Rindler, J. Dossett, J. Moldenhauer and C. Allison {\tt arXiv:0710.4726v1},  M. Sereno {\tt arXiv:0711.1802v1} and P. Bakala, P. Cermak, S. Hledik, Z. Stuchlik and K. Truparova {\tt arXiv:0709.4274v1 }
\end{thebibliography}
\end{document}